\documentclass[12pt]{iopart}

\usepackage{graphicx}
\usepackage{bbm}
\usepackage{color}

\newcommand{\msum}
{{\kern 4.5ex\raisebox{0.25ex}{$'$}\kern -4.5ex}\sum}

\newcommand{\difference}{\delta}

\renewcommand{\imath}[0]{{\rm i}}

\begin{document}

\title[Dispersion energies of BCS superconductors]{Optical BCS conductivity at 
imaginary frequencies and
dispersion energies of superconductors}

\author{G. Bimonte$^{1}$, H. Haakh$^{2}$, C. Henkel$^{2}$ 
and F. Intravaia$^{2}$}
\address{$^{1}$Dipartimento di Scienze Fisiche Universit\`{a} di Napoli Federico
II Complesso Universitario MSA,
Via Cintia, 80126 Napoli Italy and INFN Sezione di Napoli, Italy\\
$^{2}$Institut f\"{u}r Physik und Astronomie, Universit\"{a}t Potsdam,
Karl-Liebknecht-Str.\ 24/25,
14476 Potsdam, Germany}

\date{09 December 09}

\begin{abstract}
We present an efficient expression
for the analytic
continuation to arbitrary complex frequencies of the complex
optical and AC conductivity of a homogeneous superconductor with
arbitrary mean free path. Knowledge of this quantity is
fundamental in the calculation of thermodynamic potentials and
dispersion energies 
involving type-I superconducting bodies. When considered for
imaginary frequencies, our formula evaluates faster than previous
schemes involving Kramers--Kronig transforms. A number of applications
illustrates its efficiency: a simplified low-frequency expansion of the 
conductivity, the electromagnetic bulk self-energy due to longitudinal
plasma oscillations, and the Casimir free energy of a superconducting
cavity.
\end{abstract}

\maketitle

\section{Introduction}
%


The linear response of a homogeneous system to an external field
can be conveniently described by a susceptibility $\chi( {\bf q}, \omega )$
in Fourier space. Calculations at real frequencies can be directly compared
to experimental spectra. Imaginary frequencies, $\omega = {\rm i} \xi$,
are theoretically useful to deal with equilibrium thermodynamical 
properties~\cite{landau}.
In particular, quantum field theories, after a Wick rotation in imaginary time,
can be studied at nonzero temperature by expressing partition functions
as Matsubara sums over discrete frequencies 
$\omega_n = 2 \pi n {\rm i} T$ (bosons: $n = 0, 1, \ldots$; 
fermions: $n = \frac12, \frac32, \ldots$)~\cite{Matsubara55}. 
A further advantage of 
this approach comes from the fact that causal response functions are
well-behaved on the (positive) imaginary axis, a feature that facilitates
both analytical and numerical calculations~\cite{landau}.


A strong motivation for the present paper is the interest in electromagnetic
dispersion forces for superconducting materials. In the context of the 
electromagnetic Casimir effect, the role of absorption in a real mirror is still
poorly understood. The superconducting phase transition provides a clean
way to change absorption over a narrow temperature range, as has been
suggested in Ref.~\cite{Bimonte05}. The standard theory of 
Bardeen--Cooper--Schrieffer (BCS)~\cite{Schrieffer99}
is well established for conventional
superconductors and has provided the basis for calculating the optical
conductivity $\sigma( \omega )$. In their seminal paper~\cite{Mattis58},
Mattis and Bardeen included the impact of disorder in the `dirty limit'
where the scattering mean free path $\ell = v_F \tau$ is small.
This approach has been generalized to arbitrary purity~\cite{Belitz89,Zimmermann91,Berlinsky93}.



The conductivity, at imaginary frequencies, is the basic response
function that yields electromagnetic dispersion forces between 
superconducting bodies at nonzero temperature~\cite{landau,parse,Milton04}.
In previous work, the common strategy was to compute $\sigma( {\rm i} \xi )$
numerically, starting from the conductivity available at real frequencies
and performing a Kramers-Kronig transform~\cite{Bimonte05a}. 
This technique is quite
inefficient because in BCS theory, $\sigma( \omega )$ is itself an integral
over the quasiparticle spectrum. This fact also complicates
analytical studies. We present in this paper an alternative formula for
the BCS conductivity that does not
need a Kramers-Kronig transform, is given by a rapidly converging
integral, and can be evaluated in the upper half of the complex frequency 
plane. Our derivation is based on a consistent regularization scheme
where the quasiparticle response far away from the Fermi edge (that
coincides with a normal conductor) is subtracted.


The paper is organized as follows: in Sec.~\ref{MattisBardeen} we
review the BCS conductivity
at real frequencies. In Sec.
\ref{DirectCalculation} we present our formula for the BCS
conductivity $\sigma(z)$ at arbitrary complex frequencies $z$,
and we show that for $z = {\rm i} \xi$, 
it coincides with the Kramers--Kronig transform, 
evaluated numerically from
$\sigma(\omega)$ given in Ref.~\cite{Zimmermann91}.
In Sec. \ref{Applications} we discuss several applications.
First, we derive the low-frequency asymptotics and
recover in a simple way the findings of Ref.~\cite{Berlinsky93}. 
We then compare with the two-fluid model of a
superconductor~\cite{Schrieffer99} and show its limitations.
As further applications, we study the zero
temperature electromagnetic self-energy of a bulk superconductor, 
as a function of the scattering mean free path
(Sec.~\ref{Plasmon-self-energy}), and present a numerical calculation of the 
Casimir energy between two superconducting plates
near the critical temperature (Sec.~\ref{Casimir}).

Our conclusions are presented in
Section~\ref{Conclusions}. Two Appendices close the paper: in
\ref{a:normal-metal} we present an alternative computation of
the normal-metal conductivity, more in the spirit of
Ref.~\cite{Mattis58}. Finally, in App.~\ref{b:zimm}, we verify
analytically that our formula, when considered for real
frequencies, reproduces the results of Ref.~\cite{Zimmermann91}.

\vspace*{1em}

\section{Mattis-Bardeen theory}
\label{MattisBardeen}

In 1958 J. Bardeen and D.C. Mattis developed a calculation of the anomalous skin
effect
in superconductors \cite{Mattis58}. The conductivity of a bulk of
superconducting material can
be obtained from the linear current response to an applied vector potential.
Expanding the
current over the quasiparticle spectrum given by BCS theory, one finds a
Chambers-like expression for the current density \cite{Mattis58, Schrieffer99}
\begin{eqnarray}
 {\bf j}( {\bf r}, \omega ) &=& \frac{ 3 \sigma_{0} }{ (2\pi)^2 \ell }
 \int\!{\rm d}^3\!{R}
 \frac{ {\bf R} ({\bf R} \cdot {\bf A}( {\bf r} + {\bf R} ) ) }{
 R^4 } {\rm e}^{ - R / \ell } I( \omega, R)
\label{eq:current-A-nloc-relation}
\\
  I( \omega, R ) &=&
 \int\!{\rm d}\epsilon\,{\rm d}\epsilon'\left\{
 L( \omega, \epsilon, \epsilon') - \frac{f(\epsilon) - f(\epsilon')}{
 \epsilon' - \epsilon} \right\}
 \cos [ R ( \epsilon' - \epsilon ) / v_{F}]
 \label{eq:def-I-integral}
\end{eqnarray}
where $\sigma_0$ is the DC conductivity in the normal state.
Disorder is taken into account via the scattering mean free path $\ell$,
$v_F$ is the Fermi velocity,
$f( \epsilon )$ 
is the Fermi-Dirac function, and $\epsilon$ is measured from the Fermi 
edge.

The spectral function $L( \omega, \epsilon, \epsilon')$ involves
the quasiparticle energies ($\hbar = 1$)
\begin{eqnarray}
 E &=& \sqrt{ \epsilon^2 + \Delta^2 } \quad
\mbox{[$\Delta\equiv\Delta(T)$: BCS gap] }
	\label{eq:def-quasiparticle-energy}
\end{eqnarray}
and is given by
\begin{eqnarray}
 L( \omega, \epsilon, \epsilon') &=&
 {\textstyle\frac12} p(\epsilon, \epsilon')
 \left(
 \frac{ f( E' ) - f( E ) }{ E - E' - \omega - {\rm i} 0^+ }
 +
 \frac{ f( E' ) - f( E ) }{ E - E' + \omega + {\rm i} 0^+ }
 \right)
	\nonumber
	\\
	&& {}
	+
 {\textstyle\frac12}q(\epsilon, \epsilon')
 \left(
 \frac{ 1 - f( E' ) - f( E ) }{ E + E' - \omega - {\rm i} 0^+ }
 +
 \frac{ 1 - f( E' ) - f( E ) }{ E + E' + \omega + {\rm i} 0^+ }
 \right)
 \label{eq:BCS-L-function}
\end{eqnarray}
with the coherence factors
\begin{equation}
 \left. \begin{array}{c}
 p(\epsilon, \epsilon') \\
 q(\epsilon, \epsilon')
 \end{array}\right\} =
 \frac{ E E' \pm (\epsilon \epsilon' + \Delta^2) }{ 2 E E' }
 \label{eq:def-coherence-factors}
\end{equation}
The infinitesimal parameter $0^+$, set to zero at the
end of the calculation, ensures an adiabatic switching-on of the
external field with time-dependence ${\rm e}^{- {\rm i}\omega t}$ and
a causal response. The term $(f - f')/(\epsilon' - \epsilon)$ that
appears in Eq.(\ref{eq:def-I-integral})
gives an imaginary contribution
to the conductivity (for real $\omega$) and represents
the diamagnetic (London) current~\cite{Mattis58,London35}. Note that the
same expression is found by the Green function approach of
Belitz \emph{et al.} \cite{Belitz89} when the spectral strength of the material
in its
normally conducting state is considered as diffusive within a charge-conserving
approximation.

In this paper, we focus on the local limit of the current response: it
is obtained from the previous expression assuming that the
vector potential
${\bf A}( {\bf r} + {\bf R} )$ is slowly varying in space. Taking it out
of the integral and integrating over ${\bf R}$,
%
one obtains the conductivity in the form \cite{Berlinsky93}
\begin{eqnarray}
 \sigma( \omega) &=& \frac{ \sigma_0 \gamma }{ {\rm i}\omega }
 \int\!{\rm d}\epsilon\,{\rm d}\epsilon'
 \left\{
 L( \omega, \epsilon, \epsilon') - \frac{f - f'}{
 \epsilon' - \epsilon} \right\}
  D_\gamma( \epsilon - \epsilon' )
	 \label{eq:sigma-a-la-Berlinsky}
\\
 D_\gamma( x ) &:=&
 \frac{ \gamma / \pi }{ x ^2 + \gamma^2 }
	\label{eq:def-smeared-delta}
\end{eqnarray}
where the scattering rate $\gamma = 1/\tau = v_F / \ell$ 
can be interpreted as the effective width of
quasiparticle
energies at the Fermi edge [last factor in Eq.
(\ref{eq:sigma-a-la-Berlinsky})].
This expression has been further simplified by Zimmermann
 \cite{Zimmermann91} and by Berlinsky \textit{et al.}
 \cite{Berlinsky93}, but their final result was given in a form, which involves
the frequency
$\omega$ in the integration boundaries and in the argument of Heaviside
functions, so that a direct continuation to complex frequencies is
not straightforward. 
In the next Section, we show that the
analytic continuation of the BCS
conductivity~(\ref{eq:sigma-a-la-Berlinsky}) to arbitrary complex
frequencies $z$ in the upper half-plane 
$\mathbbm{C}^+$ 
 can be written in the form
\begin{equation}
 \sigma(z) = \frac{ \sigma_0 }{ 1 -{\rm i}z \tau }
 + \delta\sigma_{\rm BCS}(z)
 \label{eq:split-sigma-result}
\end{equation}
where one recognizes in the first term the analytical
continuation of the Drude result for the normal metal
$\sigma_{\rm Dr}( \omega ) = \sigma_0 / (1 - {\rm i} \omega
\tau)$. The BCS correction $\delta\sigma_{\rm BCS}(z)$ is given
by Eqs.\eref{ouran}, \eref{eq:more-regular-form-for-G} 
below that can be easily
evaluated numerically.
\section{Analytic continuation to complex frequencies}
\label{DirectCalculation}

Let us now go back to the expression given in Eq.
\eref{eq:sigma-a-la-Berlinsky}. We note that the spectral
function~(\ref{eq:BCS-L-function}) is evidently analytic for
$\omega = z \in \mathbbm{C}^+$, and the analytic continuation 
of the BCS conductivity~(\ref{eq:sigma-a-la-Berlinsky}) is immediate 
at this stage. The double integral over
$\epsilon$ and $\epsilon'$ is, however, tricky to perform. 

%

Mattis and Bardeen rewrite the integrand using symmetries under the exchange of
 $\epsilon \leftrightarrow \epsilon'$ and introduce a regulating function to
ensure the convergence at large $\epsilon$, $\epsilon'$. This makes obsolete 
the inclusion of the diamagnetic term under the integral in 
Eq.\eref{eq:sigma-a-la-Berlinsky}.
They then perform the integration along a contour in the complex
$\epsilon'$-plane, leaving only the integration over $\epsilon$ to
be done numerically. Further calculations are performed in the `dirty'
limit where the mean free path $\ell$ is much smaller than the
coherence length $v_F / \Delta$, i.e., $\gamma \gg \Delta$. 
Zimmermann~\cite{Zimmermann91} and Berlinsky \emph{et al.}~\cite{Berlinsky93}
consider a general scattering rate, but focus on the case of real frequencies.

In the following, we show that
the difficulty with the convergence can be overcome by a proper handling 
of the integration scheme: we exploit the exchange symmetry in the
$\epsilon\epsilon'$-plane and secure uniform convergence of the
double integral by a suitable subtraction. We take into account explicitly the diamagnetic current
and include a general scattering rate as in 
Refs.~\cite{Zimmermann91, Berlinsky93}.
This procedure makes the Drude term in Eq.\eref{eq:split-sigma-result} 
emerge in a natural way for complex frequencies and leads
to a numerically convenient form for the remaining BCS correction
(whose integrand is strongly localized).


\subsection{Normal metal}


%




Let us consider first the limiting case of a normal conductor,
that is the expression obtained from Eq.\eref{eq:sigma-a-la-Berlinsky} 
in the limit $\Delta \to 0$. We allow for complex $\omega$, denoted 
by $z$. 
The coherence factors in 
Eq.\eref{eq:def-coherence-factors} reduce to $\frac12[ 1 \pm
{\rm sgn}( \epsilon \epsilon') ]$. We observe that the first and
second terms of Eq.\eref{eq:BCS-L-function} are related
by the mapping $E' \leftrightarrow -E'$. It is easy to check that
in all four quadrants of the $\epsilon\epsilon'$-plane, we can
replace the spectral function
by
\begin{equation}
\Delta \to 0 :
 L(z, \epsilon, \epsilon')
 \mapsto
 \frac{ \epsilon' - \epsilon }{ (\epsilon' - \epsilon)^2 -z^2 }
 \left[ f(\epsilon) - f(\epsilon') \right]
 \label{eq:Drude-L-function-of-xi}
\end{equation}
The conductivity of the normal metal is now obtained as follows. Introducing
the variables $s = \frac12(\epsilon + \epsilon')$ and $\difference =
\epsilon' - \epsilon$, the difference of Fermi functions can be written as
[$k_B = 1$]
%
\begin{equation}
 f(\epsilon) - f(\epsilon') =
 \frac{ \sinh (\difference / 2T) }{
 \cosh (\difference / 2T) + \cosh (s / T)}
\label{eq:difference-Fermi-functions}
\end{equation}
which is integrable at $|s| \to \infty$ with the result
\begin{equation}
 \int\limits_{-\infty}^{+\infty}\!{\rm d}s
 \left[
 f(\epsilon) - f(\epsilon') \right]
 = \difference
\label{eq:sum-integral}
\end{equation}
The integration over $\difference$ is then easily performed:
\begin{equation}
 \sigma(z) =
 \frac{ \sigma_0 \gamma }{ {\rm i} z }
 \int\limits_{-\infty}^{+\infty}\!{\rm d}\difference\,
 \left\{
 \frac{ \delta^2 }{ \difference^2 -z^2 } - 1 \right\}
 D_\gamma( \difference )
  =
 \frac{ \sigma_0 \gamma }{ \gamma -{\rm i} z }
 \label{eq:Drude-result-from-BCS-Francesco}
\end{equation}
where the subtraction in curly brackets corresponds to the limit
$z \to 0$ (the diamagnetic current, compare Eqs.
(\ref{eq:sigma-a-la-Berlinsky}, \ref{eq:Drude-L-function-of-xi})).
We thus find the analytic continuation of the Drude conductivity.
\ref{a:normal-metal} presents an alternative derivation
of this result, that proceeds in close analogy to 
Ref.~\cite{Mattis58}. In this approach, the integrand is
multiplied by a factor $\exp[{\rm i}(\epsilon + \epsilon' )/(2\Lambda)]$ in
order to enforce absolute convergence in the direction $|\epsilon
+ \epsilon'| \to \infty$ and the limit $\Lambda \to \infty$ is
taken in the end.

\subsection{Superconductor}



We now return to the full BCS theory. It is a simple matter to
verify that for any complex frequency $z \in \mathbbm{C}^+$, 
the integrand involving the spectral function $L(z,\epsilon,\epsilon')$
can be split into two terms
\begin{equation} 
L(z, \epsilon, \epsilon') D_\gamma( \epsilon - \epsilon' )
= F(z, \epsilon, \epsilon') + F(z, \epsilon', \epsilon)
\;,
	\label{eq:sym}
\end{equation} 
where
\begin{eqnarray}
	F(z, \epsilon, \epsilon') &=& 
	-
	D_\gamma( \epsilon - \epsilon' )
	\frac{ \tanh( E / 2 T ) }{ 4 E }
	\left\{
	\frac{A_{+} +\epsilon \epsilon' }{ \epsilon'^2 - {Q}_{+}^2} 
	+
	\frac{A_{-} +\epsilon \epsilon' }{ \epsilon'^2 - {Q}_{-}^2} 
	\right\}
	\;,
	\label{eq:def-F-with-A-and-Q}
	\\
	A_{\pm}(z,E) &=& E(E \pm z)+\Delta^2\;,\label{Apm}
	\\
	Q^2_{\pm}(z,E) &=& (E \pm z)^2-\Delta^2\,.\label{eq:Qpm}
\end{eqnarray}
Eq.(\ref{eq:sym}) shows manifestly the symmetry of the integrand
under the exchange $\epsilon \leftrightarrow \epsilon'$. 
Mattis and Bardeen proceed by integrating separately over the two
terms $F(z, \epsilon, \epsilon')$ and $F(z, \epsilon', \epsilon)$,
but to do so, one must ensure 
absolute convergence of the double integral. Indeed, for large 
$|\epsilon|$,
$|\epsilon'|$, the $\tanh$ functions tend to $\pm 1$, and from
Eqs.(\ref{eq:def-F-with-A-and-Q}-\ref{eq:Qpm}),
we see that $F(z, \epsilon,
\epsilon')$ becomes proportional to $(\epsilon-\epsilon')^{-1}$.
Therefore for large $|\epsilon|$, $|\epsilon'|$, the integrand
$F(z, \epsilon, \epsilon')$ becomes independent of the variable 
$s = (\epsilon + \epsilon')/2$, and then its double integral is not
absolutely convergent 
as $|s| \to \infty$.

This divergence can be conveniently cured by the following
subtraction procedure. We evaluate the asymptotic limit of 
$F(z,
\epsilon, \epsilon')$ as $|\epsilon + \epsilon'| \gg \Delta$,
keeping the ratio to the other parameters $z, \,T, \, \gamma$ fixed, 
and subtract it from $F$. 
This yields a regularized integrand 
%
\begin{eqnarray}
\bar{F}(z, \epsilon, \epsilon')
&=& D_\gamma( \epsilon - \epsilon' )
\left\{ 
\frac{ - \tanh( E / 2 T ) }{ 4 E }
	\sum_{\eta = \pm}
	\frac{A_{\eta} +\epsilon \epsilon' }{ \epsilon'^2 - {Q}_{\eta}^2} 
\right.
\nonumber
\\
&& \left. +\frac12 \tanh \left( \frac{\epsilon}{2 T}\right)
\frac{ \epsilon'-\epsilon }{(\epsilon'-\epsilon)^2
-z^2}\right\}
 \,.
 \label{eq:Giuseppe-subtraction}
\end{eqnarray} 
We have checked that this difference is of order ${\cal O}[
(|\epsilon| + |\epsilon'|)^{-2} ]$ in the direction $\epsilon =
\epsilon'$ and ${\cal O}[ (|\epsilon| + |\epsilon'|)^{-5} ]$ in
all other directions of the $\epsilon\epsilon'$-plane.
In addition, there are no poles for real-valued $\epsilon$,
$\epsilon'$ so that the integral of $\bar{F}(z, \epsilon,
\epsilon')$ is uniformly convergent.

Comparison of Eqs.(\ref{eq:Drude-L-function-of-xi},
\ref{eq:Giuseppe-subtraction}) shows that
this procedure is exactly equivalent to subtracting the spectral function of
the normal conductor from the BCS expression $L(z, \epsilon, \epsilon' )$.
We hence find a form of the conductivity as
anticipated in Eq.(\ref{eq:split-sigma-result}), where the 
BCS correction is 
\begin{eqnarray}
\delta\sigma_{\rm BCS}(z)
&=&\frac{ \sigma_0 \gamma }{ {\rm i} z }
 \int\!{\rm d}\epsilon\,{\rm d}\epsilon'\,\left[ \bar{F}
 (z, \epsilon, \epsilon')+\bar{F}(z, \epsilon',
 \epsilon)\right]
 \nonumber
 \\
&=&
2\frac{ \sigma_0 \gamma }{ {\rm i} z }
 \int\!{\rm d}\epsilon\,{\rm d}\epsilon'\, \bar{F}
 (z, \epsilon, \epsilon')\;.\label{interc}
\end{eqnarray} 
In the second line, we have exploited the absolute convergence
of the integral of $\bar{F}(z, \epsilon, \epsilon')$, to interchange the
order of the $\epsilon\,,\epsilon'$ integrations for the second term.

The BCS correction $\delta\sigma_{\rm BCS}(z)$ [see 
Eq.(\ref{eq:split-sigma-result})] is now evaluated as has been done by
Mattis and Bardeen for real frequencies. We remark however that
since the subtracted integrand \textit{in toto} converges
uniformly, no cutoff is required. 
The $\epsilon'$ integration of
the quantity $\bar{F}(z, \epsilon, \epsilon')$ is easily
performed with contour techniques. We first note that thanks
to the subtraction, the integrand is at most of
order $|\epsilon'|^{-2}$ 
as $|\epsilon'| \to \infty$ at fixed $\epsilon$ so that we can close 
the contour by a semi-circle at infinity in the upper half-plane.
We then observe that 
the second term between the curly
brackets in Eq.(\ref{eq:Giuseppe-subtraction}) is
an odd function of $\epsilon'-\epsilon$ and vanishes upon
integrating over $\epsilon'$. The first term
has simple poles occurring at $\epsilon' = \epsilon + {\rm
i}\gamma$, and at $\epsilon' = Q_{\pm}$. The square
root $Q_{\pm}$ must be taken with a cut along the
real positive axis~\cite{cut-in-root},
in such a way that ${\rm Im}(Q_{\pm}) > 0$.
Evaluation of the
corresponding residues gives the result:
\begin{equation}\delta\sigma_{\rm BCS}(z)={\rm
i} \frac{\sigma_0 \,\gamma}{2 z} \int_{-\infty}^{\infty} \frac{d
\epsilon}{E} \tanh \left( \frac{E}{2 
T}\right)
\sum_{\eta=\pm} G_{\eta}(z,\epsilon)
\;,
\label{ouran}
\end{equation}
 where
\begin{equation} G_{\pm}(z,\epsilon)= \frac{{\rm i}
\gamma(A_{\pm}+\epsilon
{Q}_{\pm})}{{Q}_{\pm}[(Q_{\pm}-\epsilon)^2+ \gamma^2]}
+\frac{A_{\pm}+\epsilon(\epsilon+{\rm i} \gamma)}{(\epsilon+{\rm
i} \gamma)^2-{Q}_{\pm}^2} 
	\;.
	\label{Gpm}
\end{equation}
We note that both terms in Eq.(\ref{Gpm}) have a pole at 
$\epsilon = Q_\pm - {\rm i}\gamma$,
but the residues compensate each other exactly. This creates
numerical problems, that can be avoided by rearranging into
\begin{equation}
 G_\pm( z,\epsilon ) =
 \frac{ \epsilon^2\, Q_\pm( z, E ) + (Q_\pm( z, E ) + {\rm
i}\gamma) A_\pm( z, E ) }{Q_\pm( z, E ) [\epsilon^2 -
(Q_\pm( z, E ) + {\rm i}\gamma)^2 ]}
 \label{eq:more-regular-form-for-G}
\end{equation}
where we have restored the arguments of $Q_\pm$ and $A_\pm$
for clarity and $E = E( \epsilon )$ as in Eq.(\ref{eq:def-quasiparticle-energy}).
Note that the integrand in Eq.(\ref{ouran}) is actually even in $\epsilon$.

Eqs.(\ref{ouran}, \ref{eq:more-regular-form-for-G}), 
together with Eq.(\ref{eq:split-sigma-result}),
provide the desired analytic continuation of the BCS conductivity
to complex frequencies $z \in \mathbbm{C}^+$. In \ref{b:zimm}, we
verify that our formula, when considered for real
frequencies $\omega$, correctly reproduces the expression for
$\sigma(\omega)$ quoted in Refs.~\cite{Zimmermann91,Berlinsky93}.

Consider now the special case of a purely imaginary
frequency $z={\rm i} \xi$, that one needs for
thermal equilibrium quantities calculated in the Matsubara
imaginary-time formalism. By
using the relations $Q_-({\rm i}\xi,\epsilon)=-Q_+^*({\rm
i}\xi,-\epsilon)$, and $A_-({\rm i}\xi,\epsilon)=A_+^*({\rm
i}\xi,-\epsilon)$, it is easy to verify that the functions
$G_{\pm}({\rm i} \xi,\epsilon)$ satisfy the relation $G_-({\rm
i}\xi,\epsilon)=G_+^*({\rm i}\xi,-\epsilon)$. Then, from
Eqs.(\ref{ouran}) and (\ref{Gpm}) we obtain:
\begin{equation}
 \delta \sigma_{\rm BCS}( {\rm i}\xi ) =
 \frac{ \sigma_0 \gamma }{ \xi }
 \int\limits_{-\infty}^{+\infty}\!\frac{{\rm d}\epsilon}{E}\,
 \tanh(E / 2T) \, {\rm Re}[G_+( {\rm i}\xi,\epsilon ) ]
 \label{eq:integral-Giuseppe}
\end{equation}
In the normal conductor
limit ($\Delta \to 0$), $G_+({\rm i}\xi,\epsilon )$ becomes purely
imaginary so that Eq.(\ref{eq:integral-Giuseppe}) vanishes,
leaving only the Drude term in Eq.\eref{eq:split-sigma-result}.

From this expression we can obtain the 'extremely anomalous' or `dirty' 
limit $\gamma \gg \Delta$ considered by Mattis and
Bardeen, by expanding in powers of $1/\gamma$, and 
retaining only the lowest order:
\begin{equation}
 G_+({\rm i}\xi ,\epsilon) \approx
 \frac{{\rm i}}{\gamma} \frac{A_+({\rm i } \xi,\epsilon) }{Q_+({\rm i }
\xi,\epsilon)}~.
\label{eq:extremely-anomalous-G}
\end{equation}
In this limit, the BCS correction $\delta\sigma_{\rm BCS}$
[Eq.(\ref{eq:integral-Giuseppe})] no longer depends on $\gamma$. 
Since the approximation is valid for small frequencies $\xi
\ll \gamma$, the Drude contribution in \eref{eq:split-sigma-result}
becomes $\sigma_{\rm Dr}( {\rm i} \xi ) \approx \sigma_0$ in this limit.

\subsection{Numerical efficiency}
\label{s:KK}
%
Kramers-Kronig relations are commonly used to calculate conductivities or
dielectric functions at imaginary frequencies starting from the known representations at real frequencies (see, e.g., Ref.~\cite{Bimonte05a}).
This is particularly useful if one wants to 
extrapolate experimental data to the imaginary axis. 
The formulation we use here,
\begin{equation}
 \delta\sigma_{\rm BCS}( {\rm i}\xi ) =
 \frac{ 2 }{ \pi }
 \int\limits_0^\infty\!{\rm d}\omega
 \frac{ \omega\, {\rm Im}\, \sigma( \omega )
 }{ \omega^2 + \xi^2 }~,
 \label{eq:KK-from-imag-part-of-omega}
\end{equation}
involves the imaginary part of the conductivity
(an odd function in $\omega$). We thus avoid the
$\delta( \omega )$ singularity~\cite{Mattis58,Berlinsky93} that occurs 
in ${\rm Re}\, \sigma( \omega )$.
Eq.(\ref{eq:KK-from-imag-part-of-omega}) is numerically
quite inconvenient in the BCS theory since it requires two integrations 
(over $\epsilon$ and $\omega$).
The calculations are more easy for the
direct analytical continuation, Eqs.(\ref{eq:split-sigma-result},
\ref{eq:more-regular-form-for-G}, \ref{eq:integral-Giuseppe}),
where a single integral must be done that in addition converges rapidly.
The agreement between the two representations is excellent as one can see from
Fig.~\ref{fig:BCS-imaginary-plot}:
%
\begin{figure}
\begin{center}
\includegraphics[width=8cm]{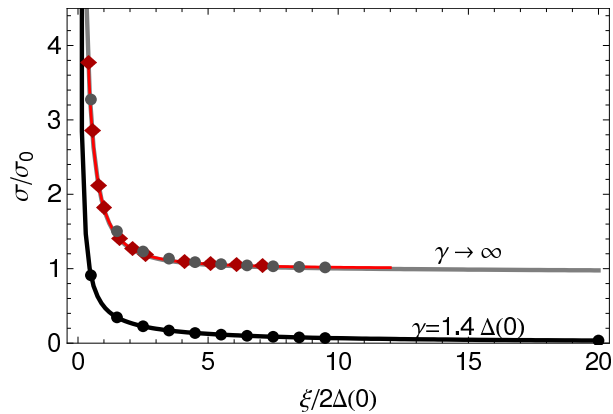}
\end{center}
\caption{(Color online)
BCS conductivity evaluated along the imaginary positive frequency axis.
Lines: Direct analytical continuation derived in Eqs.\eref{eq:split-sigma-result} 
and
(\ref{eq:more-regular-form-for-G}, \ref{eq:integral-Giuseppe}) in the clean ($\gamma
\approx 1.4\,\Delta$, black line) and dirty ($\gamma \approx
1400\,\Delta$, gray line) case and the extremely anomalous limit ($\gamma
\gg\Delta$, red line) using Eq.\eref{eq:extremely-anomalous-G}.
Dots: Kramers--Kronig continuation from the real frequency conductivities by
Zimmermann \emph{et al.} \cite{Zimmermann91}, Berlinsky \emph{et al.}
 \cite{Berlinsky93}, following the scheme summarized in 
Eq.\eref{eq:KK-from-imag-part-of-omega} for the clean and dirty case.
Diamonds: Kramers--Kronig continuation in the extremely anomalous case starting
from the complex conductivity $\sigma(\omega)$ of Ref.~\cite{Mattis58}.
}
\label{fig:BCS-imaginary-plot}
\end{figure}
%
The solid lines are obtained with our analytical continuation while the dots
result from the Kramers--Kronig scheme.
Our results also reproduce the dirty limit
based on Eq.\eref{eq:extremely-anomalous-G} ($\gamma\gg\Delta$) 
considered by
Mattis and Bardeen \cite{Mattis58} for real frequencies.
All the numerical work has been performed using the approximate form of the BCS gap function $\Delta(T)$ given in
Refs. \cite{Thouless1960, Townsend1962}.





\section{Applications}
\label{Applications}
%

\subsection{Supercurrent response}
\label{supercurrent}

As a first example, we show that the BCS response
function can be analyzed in a quite simple way at imaginary
frequencies. We recall the discussion of Berlinsky \emph{et al.}
 \cite{Berlinsky93}
who attribute the weight of the peak $\propto \delta( \omega )$ in ${\rm Re}\,
\sigma( \omega )$ to the Cooper pair condensate and the response at
nonzero frequencies to thermally excited quasiparticles
($ \omega \ll 2\Delta$) and to `direct' excitations across the gap
($\omega \ge 2\Delta$).
All three contributions together fulfill the
sum rule for the oscillator strength in the conductivity. The first and second
contributions are associated to the superfluid and the normal fluid of the
two-fluid model due to Gorter and Casimir \cite{Gorter34, Schrieffer99}, whose
weights
vary with $T$ in a way similar to the Gorter-Casimir rule. The third
contribution
is found to be approximately constant over a large range of $T$.

We have tried to evaluate the imaginary frequency representation
of these three contributions, taken separately. We recognized, however,
that the `thermal' and `direct' parts of the quasiparticle response 
[i.e., first and second term in Eq.(\ref{eq:BCS-L-function})] show
a branch cut in the $\epsilon'$-plane so that only their sum can be
combined in the same way as in Eqs.(\ref{eq:sym}--\ref{eq:Qpm}).
In what follows, we hence consider
a slightly different `two-fluid split' of the conductivity
that is closer to the Gorter and Casimir approach: a supercurrent part
that does not show any relaxation (or scattering) and a resistive current
whose frequency response is Drude-like.
In addition, we reproduce the $T$-dependent weight
of a logarithmically divergent contribution
to the conductivity at low frequencies given in Ref.~\cite{Berlinsky93}.
This behaviour is usually
attributed to the divergent mode density of the BCS approach at the gap that
is not smoothed out in the Chambers-like way disorder is introduced by Mattis
and Bardeen \cite{Mattis58}.

Along the imaginary frequency axis, the superfluid response corresponds
to the weight of the pole at $\xi = 0$. In terms of an effective superfluid plasma
frequency,
%
\begin{equation}
\xi \to 0: \qquad
 \sigma( {\rm i} \xi ) \approx \frac{ \varepsilon_0
 \omega_{\rm s}^2(T) }{ \xi } + B(T) \log\left( \Delta / \xi \right) + C(T)
 \label{eq:def-plasma-weight}
\end{equation}
where the logarithmic term is discussed below.

Comparing to Eq.\eref{eq:split-sigma-result}, we get $\omega_{\rm s}$
by considering $\delta\sigma_{\rm BCS}( {\rm i} \xi )$ in the limit
$\xi \to 0$.
Our starting point is then Eq.\eref{eq:integral-Giuseppe}.
For $\xi \to 0$, the
integrand $G_{+}({\rm i}\xi, \epsilon )$ [Eq.(\ref{eq:more-regular-form-for-G})] 
as a function of $\epsilon$ exhibits two relevant features.
First, a `narrow peak' originating from the term
$1/Q_{+}({\rm i} \xi, E(\epsilon) )$:
its width in $\epsilon$ scales like $(\xi \Delta)^{1/2}$.
Setting $\epsilon = x (\xi \Delta)^{1/2}$, we get
\begin{equation}
 \omega_{\rm s,1}^2 =
 \frac{ 2\sigma_0 \Delta }{ \varepsilon_0 } 
 \int\limits_{0}^{+\infty}\!{\rm d}x\,
 \tanh\frac{ E(\epsilon)Ê}{ 2T}
\, {\rm Re} \left[
 \frac{2 {\rm i}  }{
 \sqrt{x^2 + 2 {\rm i} }}
 \right]
	\stackrel{\xi \to 0}{\longrightarrow}
 \pi\frac{ \sigma_0 \Delta }{ \varepsilon_0 }
  \tanh( \Delta / 2 T )
 \label{eq:first-superfluid-part}
\end{equation}
where convergence at the upper limit is secured by taking
the real part. In the last step, we have taken the lowest order 
in the expansion of $E(\epsilon) = \Delta \sqrt{ 1 + (\xi / \Delta) x^2 }$ 
for small $\xi$.
In the limit $T\to 0$, we recognize here the integral term in Eq.(14) of
Berlinsky \emph{et al.}%
 \cite{Berlinsky93}. 
 
The second feature in $G_+( {\rm i}\xi, \epsilon )$
is a `broad background' whose width as a function of $\epsilon$ is
set by $\Delta$ and $\gamma$. This
gives a negative contribution to $\omega_{\rm s}^2$:
\begin{eqnarray}
 \omega_{\rm s,2}^2 &=&
 - 4 \frac{\sigma_0 \gamma }{\varepsilon_{0}} \Delta^2
 \int\limits_0^{\infty}\!{\rm d}\epsilon\,
 \frac{ \tanh( E / 2 T ) }{
 \sqrt{\Delta ^2+\epsilon ^2}
 \left(\gamma ^2+4 \epsilon ^2\right)}\nonumber
\\
 &\stackrel{T \to 0}{\longrightarrow}&
 - 4 \frac{\sigma_0}{\varepsilon_{0}} \Delta(0)
 \frac{ {\rm arcsec}(\, 2 \Delta(0) / \gamma) }{
 \sqrt{ 4 - (\gamma / \Delta(0))^2 } }
 \label{eq:Berlinsky-integral}
\end{eqnarray}
%
The zero-temperature limit given in Eq.(15) of 
Berlinsky \emph{et al.} \cite{Berlinsky93} is identical to the
sum of Eqs.(\ref{eq:first-superfluid-part},\ref{eq:Berlinsky-integral}). Recall
that in this paper, $\Delta$ always denotes the temperature-dependent
gap.

\begin{figure}[hhh]
\begin{center}
\includegraphics*[width=65mm]{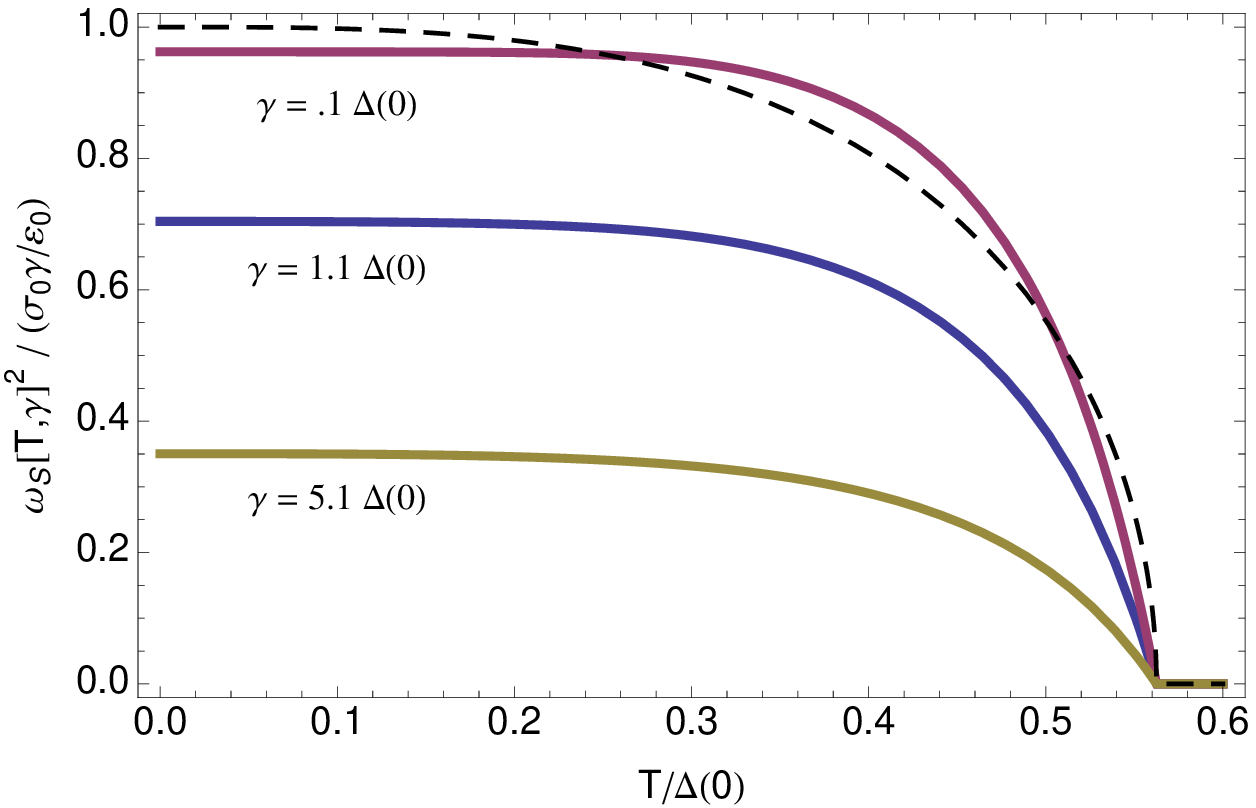}
\includegraphics*[width=67mm]{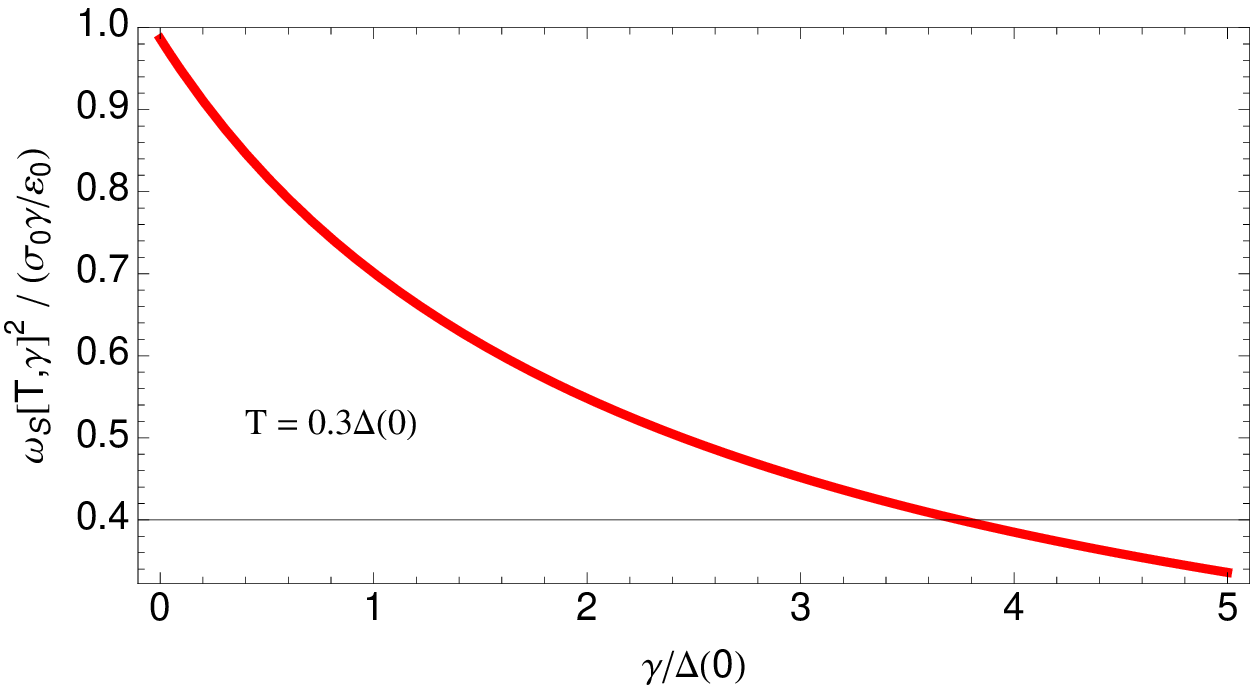}
\end{center}
\caption[]{(Color online)
Superconducting order parameter for different disorder strengths. We plot
the superfluid part of the current response function, expressed in terms
of the superfluid plasma frequency squared, $\omega_{\rm s}^2$,
and normalized to the weight of the high-frequency (above gap) response
$\omega_{\rm pl}^{2} = \sigma_{0}\gamma/\varepsilon_{0}$. This is plotted
as a function of the temperature (top) and of the scattering rate 
$\gamma$ (bottom). 
Dashed line: normalized BCS gap $\Delta( T ) / \Delta( 0 )$.
The superconducting fraction
vanishes above the critical temperature $T_{c}\approx 0.56\,\Delta(0) $ 
and becomes almost
constant for $T\ll T_{c}$. In this limit, $\omega_{\rm s}^2$ still 
depends on $\gamma$ and reaches the normal-phase plasma frequency
$\omega_{pl}^{2}$ only in the pure limit $\gamma\to 0$.}
\label{OmegaS}
\end{figure}

Fig.\ref{OmegaS} shows the behavior of $\omega_{s}^{2} =
 \omega_{\rm s,1}^2 + \omega_{\rm s,2}^2$ as a function of 
temperature and scattering rate. We normalize to the plasma frequency
given by the total carrier density,
$\omega_{\rm pl}^2 = \sigma_0 \gamma / \varepsilon_0$.
This suggests an interpretation in terms
of a superfluid order parameter~\cite{Gorter34,Bardeen1958}.
Indeed, the superfluid response identically vanishes for temperature
higher than the critical temperature
and goes to a constant in
the limit $T\ll T_{c}$. Note that at low temperatures, $\omega_{s}$ 
still depends on the disorder in the sample via the parameter $\gamma$
(see also in the following).


\subsection{Logarithmic correction}

It is well known that at nonzero temperature, the BCS
conductivity shows a logarithmically divergent term 
at low frequencies. The weight $B( T )$ of this term [see
Eq.(\ref{eq:def-plasma-weight})] can also be calculated
along imaginary frequencies, by improving on the small-$\xi$ limit
discussed before. The `narrow peak' of
Eq.(\ref{eq:first-superfluid-part}) allows to take into account the small
$\epsilon$
expansion of the temperature-dependent factor $\tanh E / 2T$.
The $\mathcal{O}( \epsilon^2 )$ term of this expansion
gives then the following integral
in the scaled energy $x = \epsilon / (\xi \Delta)^{1/2}$:
\begin{equation}
\sigma(i \xi) - \frac{\varepsilon_0 \omega^2_{\rm s}(T)}{\xi}
\approx
\frac{ \sigma_0 \Delta }{ T }
 {\rm sech}^2(\Delta / 2T)
 \int\limits_{0}^{x_c}\!{\rm d}x\,
 {\rm Re} \left[
 \frac{ {\rm i} x^2 }{
 \sqrt{x^2 + 2 {\rm i} }}
 \right]
 \label{eq:log-correction}
\end{equation}
A cutoff $x_c$ is needed for the integral in order to reproduce the convergent
behavior of the original expression. Inspection of the full integrand shows
that it starts to decrease at $\epsilon \ge \Delta$ so that we take
$x_{c} = (\Delta/\xi)^{1/2}$.
The integral can then be evaluated explicitly and gives to leading order
at small $\xi$,
$\log x_c = \frac12 \log \left( \Delta / \xi \right)$.
We thus read off the coefficient $B(T)$ in
Eq.(\ref{eq:def-plasma-weight}):
\begin{equation}
 B(T) = \frac{\sigma_0 \Delta}{2 T} {\rm sech}^2(\Delta / 2T)
 \label{eq:result-log-prefactor}
\end{equation}
It is easy to check that in the regime $\omega \ll T \ll \Delta$, our result
(analytically continued to real frequencies)
overlaps with the formulas (20, 21) of
Berlinsky \emph{et al.} \cite{Berlinsky93}.
Their result apparently allows for larger frequencies up to the order
$T$, but is restricted to $\omega, T \ll \Delta$.

\subsection{Partial sum rule}

\begin{figure}[htb]
\centering{
\includegraphics*[width=75mm]{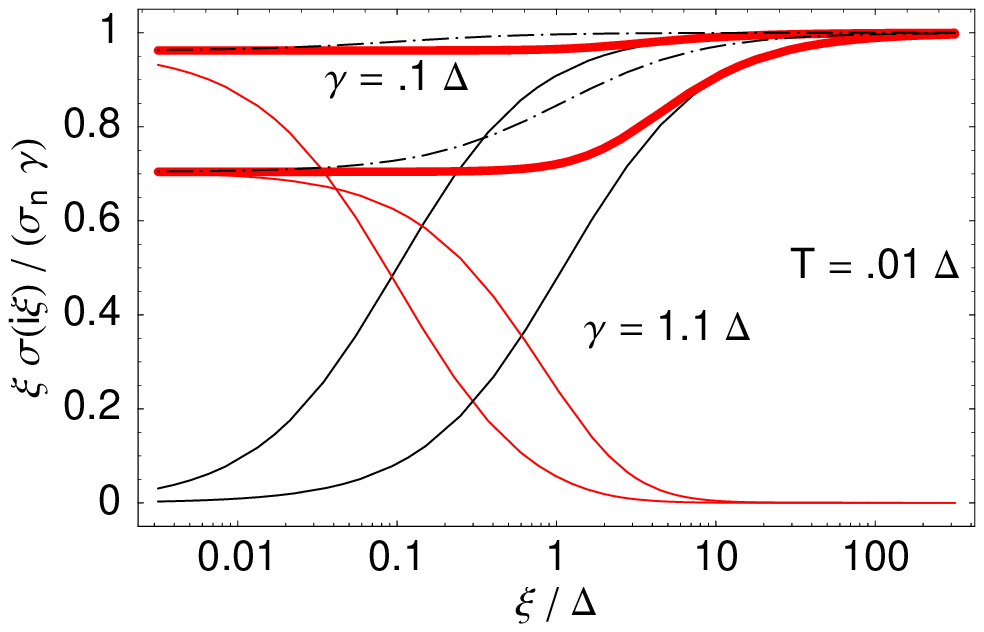}
%
\includegraphics*[width=75mm]{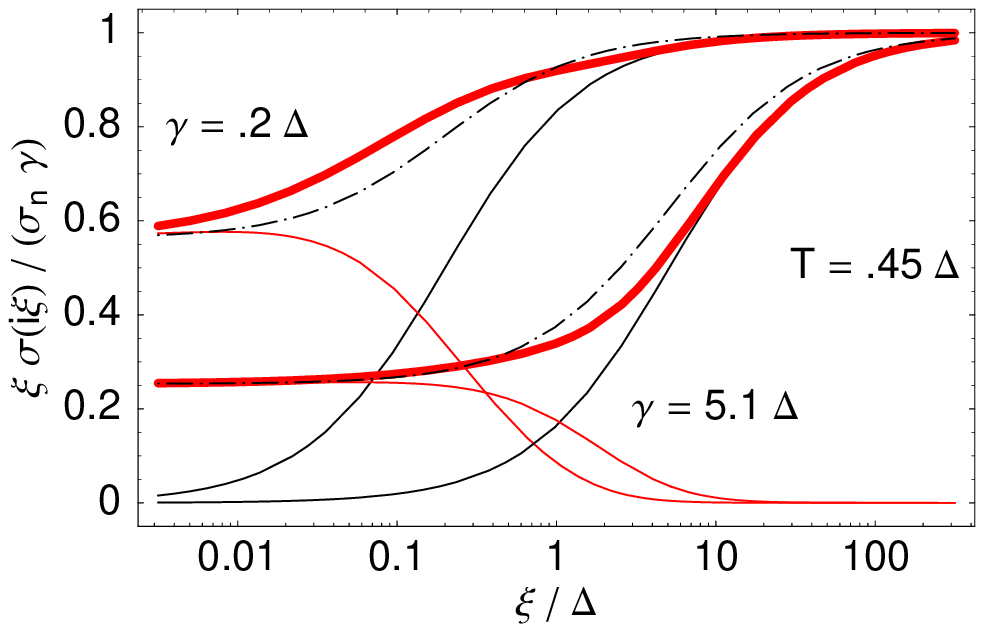}%
}
\caption[]{(Color online) Effective oscillator strength in the frequency range
$0 \le \omega \le \xi$, as measured by the ratio
$\xi \sigma( {\rm i}\xi ) / (\sigma_0 \gamma)$.
The thick red curve is based on the full BCS conductivity,
the thin black one on the Drude part only,
$\xi / ( \gamma + \xi )$,
and the thin red one on the BCS correction. Dash-dotted line: two-fluid-like
model of Eq.(\ref{eq:two-fluid-model})
with the `order parameter' $\eta(T)$ matched to reproduce the
small-$\xi$ limit.
\\
We consider a `clean' and
a `dirty' case at a low temperature (top)
and at a higher temperature (bottom). Note that $\Delta$ is always the
$T$-dependent gap here. The
shoulder at low frequencies in the `clean and warm' case
(right, top curve, $\gamma = 0.2\,\Delta$, $T = 0.45\,\Delta$)
is probably related to the logarithmic contribution in
Eq.(\ref{eq:def-plasma-weight}).}
\label{fig:sum-rule}
\end{figure}

We now make a connection between 
the conductivity at imaginary frequencies and an effective
oscillator strength, integrated over in a finite range of frequencies.
More precisely, we consider the product
$\xi \sigma( {\rm i}\xi )$
that is plotted in Fig. \ref{fig:sum-rule}.
A Kramers-Kronig relation, similar
to Eq.(\ref{eq:KK-from-imag-part-of-omega}), but based on ${\rm Re}\,
\sigma( \omega )$, shows that 
$\xi \sigma( {\rm i}\xi )$
is an non-decreasing function
of $\xi$ that collects the oscillator strength in the frequency range
$0 \le \omega \le \xi$. The sum rule for ${\rm Re}\,\sigma( \omega )$
thus implies that 
$\xi \sigma( {\rm i}\xi ) \to \gamma \sigma_0$ for $\xi \to \infty$.
According to Eq.(\ref{eq:def-plasma-weight}), the limiting value of 
$\xi \sigma( {\rm i}\xi )$
for $\xi \to 0$ is
$\varepsilon_0 \omega_{\rm s}^2(T) =: \eta( T ) \gamma \sigma_0$ 
where $0 \le \eta( T ) \le 1$ is similar to a superconducting
order parameter. A normal conductor necessarily has $\eta(T) = 0$
because its DC conductivity is finite.

We compare the product $\xi \sigma( {\rm i} \xi )$ in Fig.~\ref{fig:sum-rule}
to an effective two-fluid model with the conductivity (dash-dotted lines)
\begin{equation}
 \sigma_{\rm 2F}( {\rm i} \xi ) =
 \eta(T) \frac{ \sigma_0 \gamma }{ \xi } +
 (1 - \eta(T))\frac{ \sigma_0 \gamma }{ \xi + \gamma }
 \label{eq:two-fluid-model}
\end{equation}
where the damping rate for the normal fluid contribution (second term) is
set to $\gamma$. The overall agreement is quite good, but the logarithmic
correction is of course not captured by this approach. We also note that
the `order parameter' $\eta(T)$ introduced in this way does not 
coincide with the gap function $\Delta = \Delta( T )$, see
Eqs.(\ref{eq:first-superfluid-part}, \ref{eq:Berlinsky-integral}). In
particular, it depends both on temperature and the dissipation rate
and does not reach $\eta( T \to 0 ) = 1$ as long as $\gamma$ is nonzero (see
also Fig. \ref{OmegaS}).

\subsection{Plasmon self-energy}
\label{Plasmon-self-energy}

As a further example, we consider the electromagnetic self-energy
of a bulk superconductor at zero temperature. This quantity can be
calculated from the oscillator strength of the bulk plasmon oscillation
 \cite{Sernelius06b}.
Recall that 
plasmons are collective, longitudinal oscillations of the carrier density
that are up-shifted in energy through
the Coulomb interaction \cite{Raimes57,Ritchie57}.
We calculate here the self-energy in the local limit and find the longitudinal
(density)
response from the dielectric function
\begin{equation}
\varepsilon(\omega) =
1 + \imath\frac{\sigma(\omega)}{ \varepsilon_0 \omega}
\end{equation}
The bulk plasmon dispersion relation is indeed given by
$\varepsilon(\omega) = 0$ with solutions in the lower half-plane 
$\mathbbm{C}^-$. Along the real frequency axis, the spectral density
broadens and shifts, and applying the logarithmic argument theorem,
the self-energy is given by~\cite{Sernelius06b,Davies72,Ford85,Obcemea87}
\begin{equation}
E = 
\int_{0}^{\infty}\frac{\rm d \omega}{2\pi}\;
\frac{\omega}{2}\partial_{\omega}{\rm Im}\log \left(\frac{ 1 }{
\varepsilon(\omega) }\right)
\end{equation}
Performing a partial integration and shifting the integration path to imaginary
frequencies, we get
\begin{equation}
E =
{\rm Im}\!
\int_{0}^{\infty}\frac{\rm d \omega}{2\pi}\ 
\log \varepsilon(\omega) =
\int_{0}^{\infty}\frac{\rm d \xi}{2\pi}
\log \varepsilon(\imath \xi)
\label{eq:bulk-energy}
\end{equation}
where the $\log \xi$ divergence at $\xi \to 0$ is integrable.
For a normal conductor [$\varepsilon(\omega)$ in Drude form],
this is easily evaluated in
terms of two characteristic frequencies, $\Omega_{\rm pl}$ 
and $\gamma$:
%
\numparts
\begin{equation}
E_{D} ={\rm Re}\left[\frac{
\Omega_{\rm pl}}{2}-\imath\frac{
\Omega_{\rm pl}}{\pi}\log\frac{ \Omega_{\rm pl}}{ \gamma } \right]
,
\qquad
\Omega_{\rm pl} = \sqrt{\omega_{\rm pl}^2 - \gamma^2 / 4 }
 - \imath \gamma/2
 \label{eq:bulk-energy-drude}
\end{equation}
where $\omega_{\rm pl} = \sqrt{\gamma\sigma_{0} / \varepsilon_0}$ 
is again the plasma frequency.
The case $\gamma \ll \omega_{\rm pl}$ is the relevant one for most
materials. 
In the limit $\gamma \to 0$, we recover the zero-point
energy $
\omega_{\rm pl}/2$ for an undamped plasmon
{}[dielectric function
$\varepsilon(\omega) = 1 - \omega_{\rm pl}^{2}/\omega^{2}$].
At nonzero $\gamma$, corrections are due
to the renormalization of the real part of $\Omega_{\rm pl}$ and the logarithmic
term. The latter can be interpreted as the contribution of zero-point fluctuations 
of the bath that is responsible for the damping of the plasmon mode
\cite{Nagaev02,Intravaia08}. Alternatively, it represents how the modes of
the bath shift in frequency, due to the coupling to the plasma 
oscillator~\cite{Sernelius06b}.
We quote here only for completeness the overdamped limit
($\omega_{\rm pl}<\gamma/2$)
\begin{equation}
\label{eq:complex-plasmon-pole}
E_{D} =-\sum_{\pm}\frac{
\xi_{\pm}}{2\pi}\log\frac{ \xi_{\pm}}{ \gamma } 
,
\qquad
\xi_{\pm} =\gamma/2 \pm \sqrt{\gamma^2 / 4-\omega_{\rm pl}^2}
\:,
\end{equation}	
\endnumparts
although the Mattis-Bardeen theory for an impure superconductor~\cite{Mattis58} 
is clearly no longer valid in this limit.

\begin{figure}[tbhp]
\begin{center}
\includegraphics[width=8.5cm]{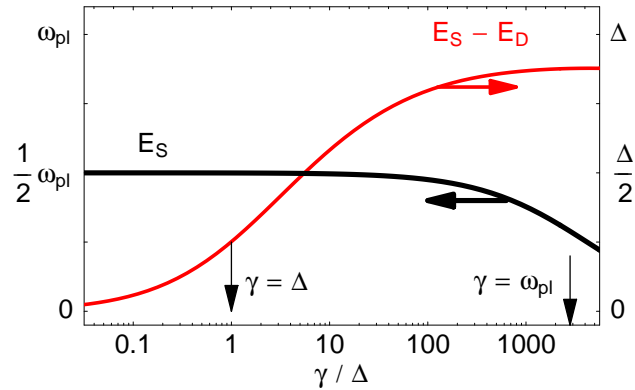}
\caption{(Color online)
Electromagnetic self-energy $E_S$ [Eq.(\ref{eq:bulk-energy}) with
the BCS dielectric function] of the bulk plasmon mode in a
superconductor vs the impurity parameter $\gamma$ (logarithmic
scale), at zero temperature.
Thick black line: self-energy $E_S$, in units of
the plasma frequency $\omega_{\rm pl} \approx 2800\,\Delta(0)$ 
(left scale). 
Red line:
difference $E_S - E_D$ to the normal conductor,
Eq.(\ref{eq:bulk-energy-drude}), in units of the (zero temperature) 
BCS gap $\Delta = \Delta(0)$ (right scale). We take for $E_D$
a Drude dielectric function with the same plasma frequency 
$\omega_{\rm pl}$ and damping rate $\gamma$.
}
\label{fig:BenG}
\end{center}
\end{figure}

For the superconductor, the self-energy $E = E_S$ follows very closely the
normal conductor, as can be seen from Fig.~\ref{fig:BenG}. The difference
arises from collective modes within the gap and modifies $E_S$ by an amount
of order $\Delta$. 
The plot features
the transition to the dirty limit~\cite{Mattis58} at
$\gamma\sim\Delta(0)$ and a saturation of $E_S - E_D$ in the overdamped
limit.

\subsection{Casimir free energy of a superconducting cavity}
\label{Casimir}
Another example where the knowledge of optical response functions at imaginary frequencies is not only highly useful, but virtually indispensable, are
electromagnetic self-energies like the Casimir interaction. 
Indeed, the calculation of these energies is practically impossible
to perform along real frequencies because of a highly oscillatory integrand.
Along the imaginary frequency axis, the integrand becomes smooth and
rapidly convergent, similar to what happens in Eq.(\ref{eq:bulk-energy}).
But the Casimir effect in
a superconducting cavity is very promising also from
a physical viewpoint because the superconductor offers 
the possibility to control dissipation. The influence of the latter
on the Casimir free energy 
remains indeed an open 
question~\cite{Milton04, Klimchitskaya06b,Brown-Hayes06}:
two conflicting viewpoints have been raised as to whether the DC conductivity
of the mirrors should be included in the modelling or not. 
It has therefore been proposed to resolve this issue by considering 
the Casimir energy of a superconducting
cavity, more precisely its change across the critical temperature
\cite{Bimonte05,  Bimonte05a, Bimonte08}. 
Earlier calculations in the framework of BCS theory~\cite{Bimonte05a} had to use 
a Kramers-Kronig analytic continuation of the optical response functions to 
imaginary frequencies [Eq.(\ref{eq:KK-from-imag-part-of-omega})], leading to
a significant computational overhead. 
The scheme developed in this work involves requires less integrations and 
enables us to perform numerical analysis in a much more flexible and precise way.

We recall that the Casimir free energy per unit area  \cite{parse} 
between two plates (separation $L$, temperature $T$) 
is given by a sum over the Matsubara frequencies 
$\xi_n = 2 \pi n T$, 
\begin{eqnarray}
\label{eq:casimir_energy}
\mathcal{F} (L,T)&=& \frac{
T}{2 \pi} \msum \int_0^\infty k dk \sum_{p} \log D_p(i \xi_n, k)\\
  D_p (\omega, k)&=&1 - r_p^2(\omega, k) \exp\left(- 2 \kappa L \right)~,\nonumber
\end{eqnarray}
where the primed sum weighs the zeroth term by $1/2$ and the last sum is over polarizations $p \in \{\rm TE, \rm TM\}$. The simplest model for the
reflection amplitudes are the Fresnel equations
\begin{equation}
r_{\rm TE}(\omega, k)=\frac{\kappa-\kappa_{m}}{\kappa+\kappa_{m}},
\quad
r_{\rm TM}(\omega, k)=\frac{\epsilon(\omega)\kappa-\kappa_{m}}{\epsilon(\omega)\kappa+\kappa_{m}}~
\label{eq:fresnel}
\end{equation}
with the notation $\kappa = \sqrt{k^2 -\frac{\omega^2}{c^2}}$ 
and  $\kappa_m = \sqrt{k^2 -\epsilon(\omega)\frac{\omega^2}{c^2}}$
(for $\omega \in \mathbbm{C}^+$, 
${\rm Re}\, \kappa \ge 0$ and
${\rm Im}\, \kappa \le 0$).
In the numerical calculation of the Matsubara sum from Eq.\eref{eq:casimir_energy}, the sum has been cut off at a sufficiently 
high~\cite{Haakh09} value of $n$
and the zeroth summand ($\xi_0 =0$) has been evaluated using the results of Sec. \ref{supercurrent}.

\begin{figure}[hhh]
\centering{
\includegraphics*[width=85mm]{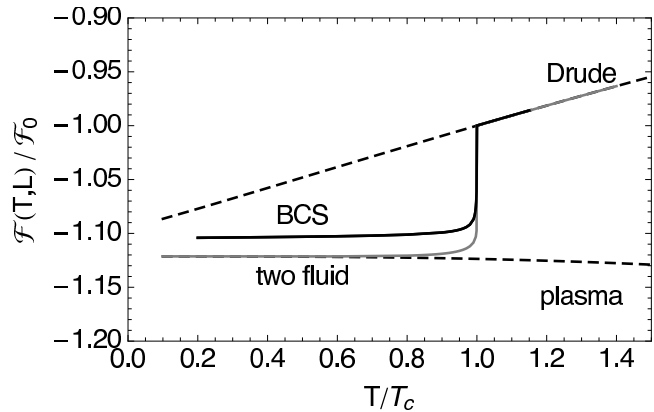}
}
\caption[]{Casimir free energy (per unit area) vs. temperature in a planar
cavity of length $L = 100\,\lambda_{\rm p}$ 
where $\lambda_{\rm p} = 2\pi c / \omega_{\rm pl}$ is the plasma 
penetration depth.
For numerical work we used values for niobum \cite{Romaniello06} 
$\omega_{\rm pl} =1.408 \times10^{16} {\rm rad/s}$, 
$\gamma =  2.44\times10^{14}{\rm rad/s}$, 
$T_c = 9.25\, {\rm K}$, 
and 
$\Delta( 0 ) = 1.7\, T_c 
\approx 8.5 \times10^{-4} \gamma$ (dirty limit).
Energies are normalized to $\mathcal{F}_0 \approx 0.2\,T_c / L^2$,
the absolute value of the free energy 
in the Drude model at $T_c$.
The Matsubara sum~(\ref{eq:casimir_energy}) 
is computed including up to 10,000 terms at the
lowest temperatures.
}
\label{fig:casimir}
\end{figure}

Fig. \ref{fig:casimir} shows the temperature dependence of the Casimir free 
energy calculated for different theoretical descriptions using the parameters for niobium.
Both the BCS description and the simple two-fluid model 
[Eq.(\ref{eq:two-fluid-model}) with 
$\eta(T) = 1-(T/T_c)^4$] predict a sudden change of the 
Casimir energy across the superconducting 
transition. 
As estimated in Refs.~\cite{Bimonte05,  Bimonte05a, Bimonte08},
this effect should be observable in experiments.

Having in mind the discussion from Sec.\ref{supercurrent}, it is not surprising that there is an offset between the BCS curve and the two-fluid model (or plasma) curve at low temperatures. This is because impurity scattering 
renormalizes the supercurrent plasma frequency $\omega_s$, leading to
a smaller effective oscillator strength in the BCS conductivity,
in particular for low $\xi$.

\section{Conclusions}
\label{Conclusions}
In this article we have analyzed the analytical
continuation of the bulk BCS conductivity to imaginary frequencies. 
The final result is numerically more
efficient than previously used
schemes based on Kramers--Kronig relations. Our approach simplifies the
direct calculation
of most of the involved integrals, and displays clearly convergence issues.
It also illustrates how the conductivity of the \textit{normal} component
emerges naturally from the BCS framework. The superconducting correction 
vanishes as the BCS gap $\Delta\to 0$ and is given in terms of a rapidly
convergent integral. 
The result is convenient for both analytical and numerical computations of the
optical response of superconductors. We have illustrated this by re-deriving
in a simple way
the weight of the supercurrent response, including its dependence on
temperature $T$ and the mean free path $\ell = v_F / \gamma$, and the
weight of the logarithmically divergent term in the conductivity. As another
illustration, we have given numerical calculations of the change in the
bulk plasmon self-energy at zero temperature, showing how impurity
scattering increases the self-energy, although the superconducting gap
lies well below the bulk plasmon resonance. Finally we have demonstrated how the Casimir energy for the case of a cavity made of superconducting material can be calculated numerically in an efficient way and recovered the energy change in the superconducting transition.

A more detailed investigation including the discussion of the distance dependence of the free energy and possibilities to tune Casimir energies through the dissipation rate $\gamma$ will be presented elsewhere.
Future work will also discuss the temperature-dependence of the self-energy.
Preliminary studies have revealed features (related to the behaviour of
$\sigma( {\rm i}\xi )$ near $\xi = 0$) that lead to similar physical effects
as in the thermal correction to the Casimir energy
between metals or ideal conductors~\cite{Milton04, Klimchitskaya06b,Brown-Hayes06}.
It would be also interesting to extend the calculations beyond
the local (or macroscopic) limit, taking into account a finite wave-vector
${\bf q}$, as done by P\"opel \cite{Poepel89}. This would also provide a 
natural scheme for calculating the electromagnetic self-energy of a
superconducting surface \cite{Barton79,Horing85}.

\section{Acknowledgments}
The work of FI was financially supported by the Alexander von Humboldt Foundation. 
We acknowledge funding from the ESF research network programme 
``New Trends and Applications of the Casimir Effect'',
the Deutsche Forschungsgemeinschaft, and the German-Israeli Foundation
for Scientific Research and Development.

\appendix

\section{Alternative calculation for the normal metal}
\label{a:normal-metal}
This approach is closer to the way in which Mattis and Bardeen performed the
integration.
We start
by combining Eqs.(\ref{eq:sigma-a-la-Berlinsky},
\ref{eq:Drude-L-function-of-xi})
and get for a complex frequency $z \in \mathbbm{C}^+$:
\begin{eqnarray}
 \tilde\sigma( z ) &=& \frac{ \sigma_0 \gamma }{ 2 {\rm i} z }
 \int\!{\rm d}\epsilon\,{\rm d}\epsilon'
 \left[ \tanh(\epsilon'/2T) - \tanh(\epsilon/2T) \right]
\nonumber
 \\
&& {} \times
 \frac{ \epsilon' - \epsilon }{ (\epsilon' - \epsilon)^2 - z^2 }
D_\gamma( \epsilon - \epsilon' )
 \, {\rm e}^{{\rm i}(\epsilon + \epsilon')/2\Lambda} 
 \label{eq:sigma-a-la-Berlinsky-app}
\end{eqnarray}
where the limit $\xi \to 0$ of the integral has still to be subtracted.
We have introduced a convergence factor
${\rm e}^{{\rm i}(\epsilon + \epsilon')/2\Lambda}$
which ensures absolute convergence
also in the sum variable $\epsilon + \epsilon'$, when analytically
continued into the upper half plane. The order of
integrations is now immaterial.
Following Mattis and Bardeen, we can
handle separately the two terms with the
hyperbolic functions:
integrate the term proportional to $\tanh(\epsilon'/2T)$
over $\epsilon$ (closing the contour in the upper half of the complex
$\epsilon$-plane) and over $\epsilon'$
the term involving $\tanh(\epsilon/2T)$. Summing the two results, 
this gives
\begin{equation}
 \tilde\sigma( z ) = \frac{ \sigma_0 \gamma }{ \pi {\rm i} z }
  \int\!{\rm d}\epsilon
  \tanh( \epsilon / 2T )
  \frac{ \pi {\rm i} \gamma \, {\rm e}^{ {\rm i} \epsilon / \Lambda } }
   { z^2 + \gamma^2 }
  \left( {\rm e}^{ - \gamma / 2 \Lambda } -
  {\rm e}^{ {\rm i} z / 2\Lambda } \right)
 \label{eq:MB-trick-1}
\end{equation}
Due to the convergence factor, the integral over $\epsilon$ does not vanish
by parity. It can be calculated by considering a rectangular contour in
the complex $\epsilon$-plane that encloses a strip of height $2\pi T$
in the upper half-plane. In this way, we find
\begin{equation}
 \tilde\sigma( z ) = \frac{ \sigma_0 \gamma }{ \pi {\rm i} z }
 \frac{ 2\pi {\rm i} T }{ \sinh( \pi T / \Lambda ) }
  \frac{ \pi {\rm i} \gamma \, }
   { z^2 + \gamma^2 }
  \left( {\rm e}^{ - \gamma / 2 \Lambda } -
  {\rm e}^{ {\rm i} z / 2\Lambda } \right)
  \to
  \frac{ \sigma_0 \gamma^2 }{ {\rm i} z ( \gamma - {\rm i} z ) }
 \label{eq:MB-trick-2}
\end{equation}
where the limit $\Lambda \to \infty$ was taken in the last step. We still have
to subtract the limiting case $z \to 0$, i.e.\ the diamagnetic term. This
gives the Drude conductivity $\sigma_{\rm Dr}( z )$ of
Eq.(\ref{eq:Drude-result-from-BCS-Francesco}).

\section{Comparison with Zimmermann's formula}
\label{b:zimm}
%
In this Appendix, we prove that our formula for the conductivity,
Eqs.(\ref{eq:split-sigma-result}), (\ref{ouran}) and (\ref{Gpm}),
when considered for real (positive) frequencies $\omega$ is
equivalent to the following expression, quoted in Ref.~\cite{Zimmermann91}, and reproduced here for the convenience of
the reader (we take $\hbar = 1$): 
\begin{equation} \sigma(\omega)={\rm
i}\frac{\sigma_0 \gamma}{2 \omega} \left(J+\int_{\Delta}^{\infty}
d E \,I_2\right)\;,\label{zimm1}
\end{equation} 
\begin{eqnarray} J(\omega \le 2
\Delta) &=& \int_{\Delta}^{\omega + \Delta}dE\,I_1\;,
\\
J(\omega
\ge 2 \Delta)&=&\int_{\Delta}^{\omega-\Delta} d E
I_3+\int_{\omega-\Delta}^{\omega+\Delta}d E I_1\,,
\end{eqnarray}
\begin{equation} 
I_1 = \tanh
\left(\frac{E}{2 T}\right)\left\{\left[1-\frac{A_-}{P_4
P_2}\right]\frac{1}{P_4+P_2+{\rm i} \gamma}
-\left[1+\frac{A_-}{P_4 P_2}\right]\frac{1}{P_4-P_2+{\rm i}
\gamma}\right\}\;,
\end{equation}
\begin{equation}
\eqalign{
I_2 &= \tanh
\left(\frac{E+\omega}{2 T}\right)\left\{\left[1+\frac{A_+}{P_1
P_2}\right]\frac{1}{P_1-P_2+{\rm i} \gamma}
-\left[1-\frac{A_+}{P_1 P_2}\right]\frac{1}{-P_1-P_2+{\rm i}
\gamma}\right\} 
\\
& {} +\tanh
\left(\frac{E}{2 T}\right)\left\{\left[1-\frac{A_+}{P_1
P_2}\right]\frac{1}{P_1+P_2+{\rm i} \gamma}
-\left[1+\frac{A_+}{P_1 P_2}\right]\frac{1}{P_1-P_2+{\rm i}
\gamma}\right\}\;,
}\end{equation}
\begin{equation}
I_3 = \tanh
\left(\frac{E}{2 T}\right)\left\{\left[1-\frac{A_-}{P_3
P_2}\right]\frac{1}{P_3+P_2+{\rm i} \gamma}
-\left[1+\frac{A_-}{P_3 P_2}\right]\frac{1}{P_3-P_2+{\rm i}
\gamma}\right\}\;,
\end{equation}
 where 
\begin{equation}
P_1=\sqrt{(E+\omega)^2-\Delta^2}\,\;\;\;,\;\;P_2=\sqrt{E^2-\Delta^2}
\end{equation} 
\begin{equation}
P_3=\sqrt{(E-\omega)^2-\Delta^2}\,\;\;\;,\;\;P_4={\rm i}\sqrt{\Delta^2-(E-\omega)^2}\;.\label{zimm2}
\end{equation}
In the relevant
ranges of $\omega$, the arguments of the square roots are
non-negative real numbers, and the square roots are then defined
to be positive. This conductivity differs from our Eqs.
(\ref{eq:split-sigma-result}), (\ref{ouran}) 
and (\ref{eq:more-regular-form-for-G})
by the presence of the frequency $\omega$ in some of the integration
boundaries. To demonstrate the equivalence with our formulae,
we show as a first step that Eqs.(\ref{zimm1}-\ref{zimm2}) 
can be split into a Drude term and a BCS correction, as in
Eq.(\ref{eq:split-sigma-result}). We note
that the quantities $P_1$, $P_3$
and $P_4$ above are related to our quantities $Q_{\pm}(\omega,E)$
as follows:
%
\begin{equation} P_1(\omega, E)=Q_+( \omega +{\rm
i}\,0^+,E)\;,\label{rel1}
\end{equation}
\begin{equation}
P_3(\omega \ge 2 \Delta ,E)={ Q}_-( \omega+{\rm i}\,0^+,E)
\;\;\;\;\;\; {\rm for}\;\Delta \le E \le \omega - \Delta
\end{equation}
%
and
\begin{equation} P_4( \omega\le 2 \Delta,
E)={Q}_-(
\omega+{\rm i}\,0^+,E)\;\;\;{\rm for}\;\Delta \le E \le 
\omega - \Delta .\label{rel3}
\end{equation}
\begin{equation} P_4( \omega \ge 2
\Delta,
E)={Q}_-( \omega+{\rm i}\,0^+,E)\;\;\;{\rm for}\;\omega-\Delta \le
E \le \omega + \Delta .\label{rel4}
\end{equation}
%
Using relations (\ref{rel1}-\ref{rel4}) above, we
verify that 
for all non-negative real frequencies $\omega$,
Eqs.(\ref{zimm1}-\ref{zimm2})
can be recast in the following compact form
\begin{equation}
\sigma(\omega)=\frac{{\rm i} \sigma_0 \gamma }{2 \omega } 
H( \omega)+ \delta \sigma(\omega)\;,\label{splitzim}
\end{equation} 
where
\begin{eqnarray}
H(\omega) &=& \int_{\Delta}^{\infty}\!\!
dE\,\left[\tanh\left(\frac{E+\omega}{2 T}\right) B _-(\omega,E+\omega) 
\right.
\nonumber
\\
&& {} 
\left.
-\theta(E-\Delta-\omega)\tanh\left(\frac{E}{2 T}\right) 
B_-(\omega,E)\right]
\label{anom}\end{eqnarray}
and
\begin{equation} \delta
\sigma(\omega)=
\int_{\Delta}^{\infty}\!\! dE\,\tanh\left(\frac{E}{2
T}\right)\;\sum_{\alpha=\pm}B_{\alpha}(\omega, E)
 \;,\label{zimmdelsig}
\end{equation}
 with
\begin{equation}
B_{\pm}(\omega,E)=\left(1-\frac{A_{\pm}}{Q_{\pm}
P_2}\right)\frac{1}{Q_{\pm}+P_2+{\rm i} \gamma}
-
\left(1+\frac{A_{\pm}}{Q_{\pm}
P_2}\right)\frac{1}{Q_{\pm}-P_2+{\rm i} \gamma}\;.\label{B}
\end{equation}
We can now show that the first term on the r.h.s. of Eq.(\ref{splitzim}) coincides with the normal metal contribution.
For this purpose, consider the
quantity $H(\omega)$. If the integral of $\tanh(E/2T) B_-(\omega,E)$ were
absolutely convergent at infinity, it would be possible to perform
the shift $E+\omega \rightarrow E$ in the integration variable $E$ in
the first term between the square brackets of Eq.(\ref{anom}),
and then we would find that $H(\omega)$ is zero. However, since for
large values of $E$, the quantity $\tanh(E/2T)B_-(\omega,E)$ has the
asymptotic expansion
\begin{equation} \tanh(E/2T)\,B_-(\omega,E)=\frac{2}{\omega+{\rm i}
\gamma}+\mathcal{O}(E^{-2})\;,\label{asexp}
\end{equation}
its integral diverges at
infinity. Therefore the shift of integration variable is not
permitted, and the conclusion that $H(\omega)$ vanishes is not
warranted. To evaluate $H(\omega)$, consider the following cut-off
function $f_{\Lambda}(E)$: 
\begin{equation}
f_{\Lambda}(E)=\frac{\Lambda^2}{E^2+\Lambda^2}\;.
\end{equation}
Thanks to the difference appearing in the square brackets of 
Eq.(\ref{anom}), the integral over $E$ is absolutely convergent 
at infinity, we have the identity:
$$ H(\omega)=\lim_{\Lambda \rightarrow \infty }\int_{\Delta}^{\infty}\!\!
dE\,\left[\tanh\left(\frac{E+\omega}{2 T}\right) B _-(\omega,E+\omega)
\right.$$
\begin{equation}
 \left.-\theta(E-\Delta-\omega)\tanh\left(\frac{E}{2
T}\right) B _-(\omega,E)\right]\,f_{\Lambda}(E)\;.\label{anombis}
\end{equation}
Since, for any $\Lambda < \infty$, the cut-off integrals of the two terms
between the square brackets are individually convergent at
infinity, it is now legitimate to shift the integration variable
in the first term. After we do that, we get:
\begin{equation}
H(\omega)=\lim_{\Lambda
\rightarrow \infty }\int_{\Delta+\omega}^{\infty}\!\! dE\,
\tanh\left(\frac{E}{2 T}\right) B _-(\omega,E)
[f_{\Lambda}(E-\omega)-f_{\Lambda}(E)]\;.\label{anomter}
\end{equation}
For large values of $\Lambda$ the difference between the cut-off functions
is appreciably different from zero only for large values of $E$,
and therefore, we can replace $\tanh(E/2T)B_-(\omega,E)$ by its
large-$E$ expansion, Eq.(\ref{asexp}). Inserting
into Eq.(\ref{anomter}), the evaluation of the elementary
$E$-integral results into: 
\begin{equation}
H(\omega)=\frac{2 \omega}{\omega+{\rm i}\gamma}\;
\end{equation}
After substituting this expression for $H(\omega)$ into
Eq.(\ref{splitzim}), we indeed find as promised
that its first term reproduces the normal-metal conductivity
$\sigma_0 \gamma /(\gamma - {\rm i}\omega)$. This shows that Eqs.
(\ref{zimm1}-\ref{zimm2}) for the BCS conductivity
$\sigma(\omega)$, once recast in the form of Eq.(\ref{splitzim}),
are indeed analogous to our split form 
Eq.(\ref{eq:split-sigma-result}) of the conductivity. 

What then
remains to show is that our expression for $\delta \sigma_{\rm
BCS}(z)$, given in Eqs.(\ref{ouran}), (\ref{eq:more-regular-form-for-G}), 
when considered for real frequencies, coincides with the quantity
$\delta \sigma(\omega)$ in Eqs.(\ref{zimmdelsig}), (\ref{B}).
This can be worked our in a straightforward way by making the
change of variables $\epsilon \rightarrow {\rm sgn} (\epsilon) E$
in Eq.(\ref{ouran}). We omit details of this lengthy calculation
for brevity.

%

\begin{thebibliography}{10}
\bibitem{cut-in-root}
In terms of a square root $\mathtt{sqrt}$ whose cut is along the
negative real axis, $Q_\pm = {\rm i} \,\mathtt{sqrt}[ \Delta^2 - (E \pm z)^2]$.
\end{thebibliography}


\end{document}